\documentclass[letter]{aa}  

\usepackage{subfig}
\usepackage{graphicx}
\usepackage{caption}

\usepackage{natbib}
\usepackage{hyperref}

\usepackage{txfonts}
\usepackage{times}
\usepackage{xspace}
\usepackage{gensymb} 
\usepackage{amsmath}
\usepackage{amssymb}

\usepackage{comment}
\usepackage{xcolor}

\newcommand{\Msun}{\ensuremath{\mathrm{M}_\odot}\xspace}

\newcommand{\chieff}{\ensuremath{\chi_\mathrm{eff}}\xspace}
\newcommand{\mchirp}{\ensuremath{\mathrm{M}_\mathrm{chirp}}\xspace}
\newcommand{\rdunits}{\ensuremath{\mathrm{Gpc}^{-3}\,\mathrm{yr}^{-1}}\xspace}

\usepackage{soul}

\begin{document}

   \title{Probing the progenitors of spinning binary black-hole mergers with long gamma-ray bursts}
   \titlerunning{Probing the progenitors of spinning BBH mergers with LGRBs}
   \authorrunning{Bavera et al.}
   \author{Simone\,S.\,Bavera,\inst{1}\fnmsep\thanks{E-mail: simone.bavera@unige.ch}
            Tassos\,Fragos,\inst{1}
            Emmanouil\,Zapartas,\inst{1}
            Enrico\,Ramirez-Ruiz,\inst{2,3}
            Pablo\,Marchant,\inst{4}
            Luke\,Z.\,Kelley,\inst{5}
            Michael\,Zevin,\inst{6}
            Jeff\,J.\,Andrews,\inst{5}
            Scott\,Coughlin,\inst{5}
            Aaron\,Dotter,\inst{5}
            Konstantinos\,Kovlakas,\inst{1}
            Devina\,Misra,\inst{1}
            Juan\,G.\,Serra-Perez,\inst{5}
            Ying\,Qin,\inst{5,7}
            Kyle\,A.\,Rocha,\inst{5}
            Jaime\,Román-Garza,\inst{1}
            Nam\,H.\,Tran\inst{3} \and
            Zepei\,Xing\inst{1}
          }

   \institute{
   Département d’Astronomie, Université de Genève, Chemin Pegasi 51, CH-1290 Versoix, Switzerland
   \and
Department of Astronomy and Astrophysics, University of California, Santa Cruz, CA 95064, USA
   \and
DARK, Niels Bohr Institute, University of Copenhagen, Jagtvej 128, 2200 Copenhagen, Denmark
   \and
Institute of Astrophysics, KU Leuven, Celestijnenlaan 200D, 3001, Leuven, Belgium
   \and
Center for Interdisciplinary Exploration and Research in Astrophysics (CIERA) and Department of Physics and Astronomy, Northwestern University, 1800 Sherman Avenue, Evanston, IL 60201, USA 
   \and
Enrico Fermi Institute and Kavli Institute for Cosmological Physics, The University of Chicago, 5640 South Ellis Avenue, Chicago, Illinois 60637, USA 
   \and
Department of Physics, Anhui Normal University, Wuhu, Anhui 241000, China
             }

   \date{Accepted December 03, 2021}

 
  \abstract{

    Long-duration gamma-ray bursts are thought to be associated with the core-collapse of massive, rapidly spinning stars and the formation of black holes. However, efficient angular momentum transport in stellar interiors, currently supported by asteroseismic and gravitational-wave constraints, leads to predominantly slowly-spinning stellar cores. Here, we report on binary stellar evolution and population synthesis calculations, showing that tidal interactions in close binaries not only can explain the observed sub-population of spinning, merging binary black holes but also lead to long gamma-ray bursts at the time of black-hole formation. Given our model calibration against the distribution of isotropic-equivalent energies of luminous long gamma-ray bursts, we find that $\approx$10\% of the GWTC-2 reported binary black holes had a luminous long gamma-ray burst associated with their formation, with GW190517 and GW190719 having a probability of $\approx$85\% and $\approx$60\%, respectively, being among them. Moreover, given an assumption about their average beaming fraction, our model predicts the rate density of long gamma-ray bursts, as a function of redshift, originating from this channel. For a constant beaming fraction $f_\mathrm{B}\sim 0.05$ our model predicts a rate density comparable to the observed one, throughout the redshift range, while, at redshift $z \in [0,2.5]$, a tentative comparison with the metallicity distribution of observed LGRB host galaxies implies that between 20\% to 85\% of the observed long gamma-ray bursts may originate from progenitors of merging binary black holes. The proposed link between a potentially significant fraction of observed, luminous long gamma-ray bursts and the progenitors of spinning binary black-hole mergers allows us to probe the latter well outside the horizon of current-generation gravitational wave observatories, and out to cosmological distances.
  
  }


   \maketitle
%

\section{Introduction}

The substantial increase in the sample size of merging binary black holes (BBHs) detected by the Advanced LIGO \citep{2015CQGra..32g4001L} and Advanced Virgo \citep{2015CQGra..32b4001A} detectors has allowed for significant improvement in our understanding of BBH assembly, primarily driven by meaningful population inferences.
The second gravitational-wave transient catalog, GWTC-2, contains 46 confident BBH detections \citep{2020arXiv201014527A}. 
Each system can be characterised by the chirp mass \mchirp and the effective spin parameter \chieff. Here, $\mchirp=(m_1 m_2)^{3/5}/(m_1+m_2)^{1/5}$ where $m_{1}$ and $m_{2}$ are the BH masses and $\chieff = (m_1 \textbf{a}_1 + m_2 \textbf{a}_2)/(m_1 + m_2) \cdot \hat{\textbf{L}}$ where $\textbf{a}_1$ and $\textbf{a}_2$ the BH dimensionless spin vectors and $\hat{\textbf{L}}$ the orbital angular momentum (AM) unit vector.
The majority of the detected BBHs have a $\chieff$ consistent with zero, 9 events have positive $\chieff$ at 95\% credibility, while no individual BBH events are observed with confidently negative $\chieff$. 
These observations indicate the existence of a sub-population of spinning BBHs. 

\begin{figure*}
\centering
\subfloat{\includegraphics[width=0.5\linewidth]{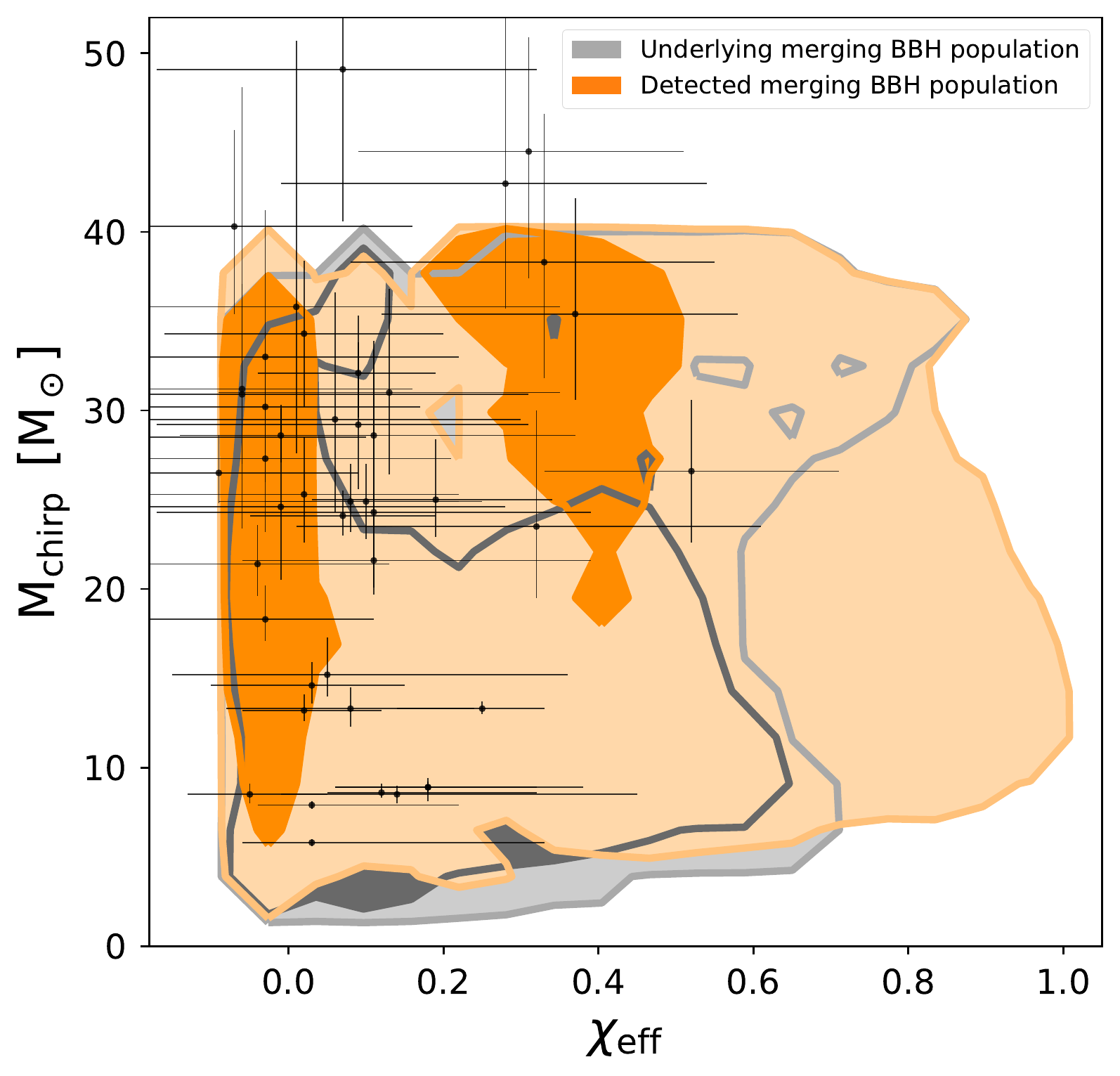}}%
\subfloat{\includegraphics[width=0.5\linewidth]{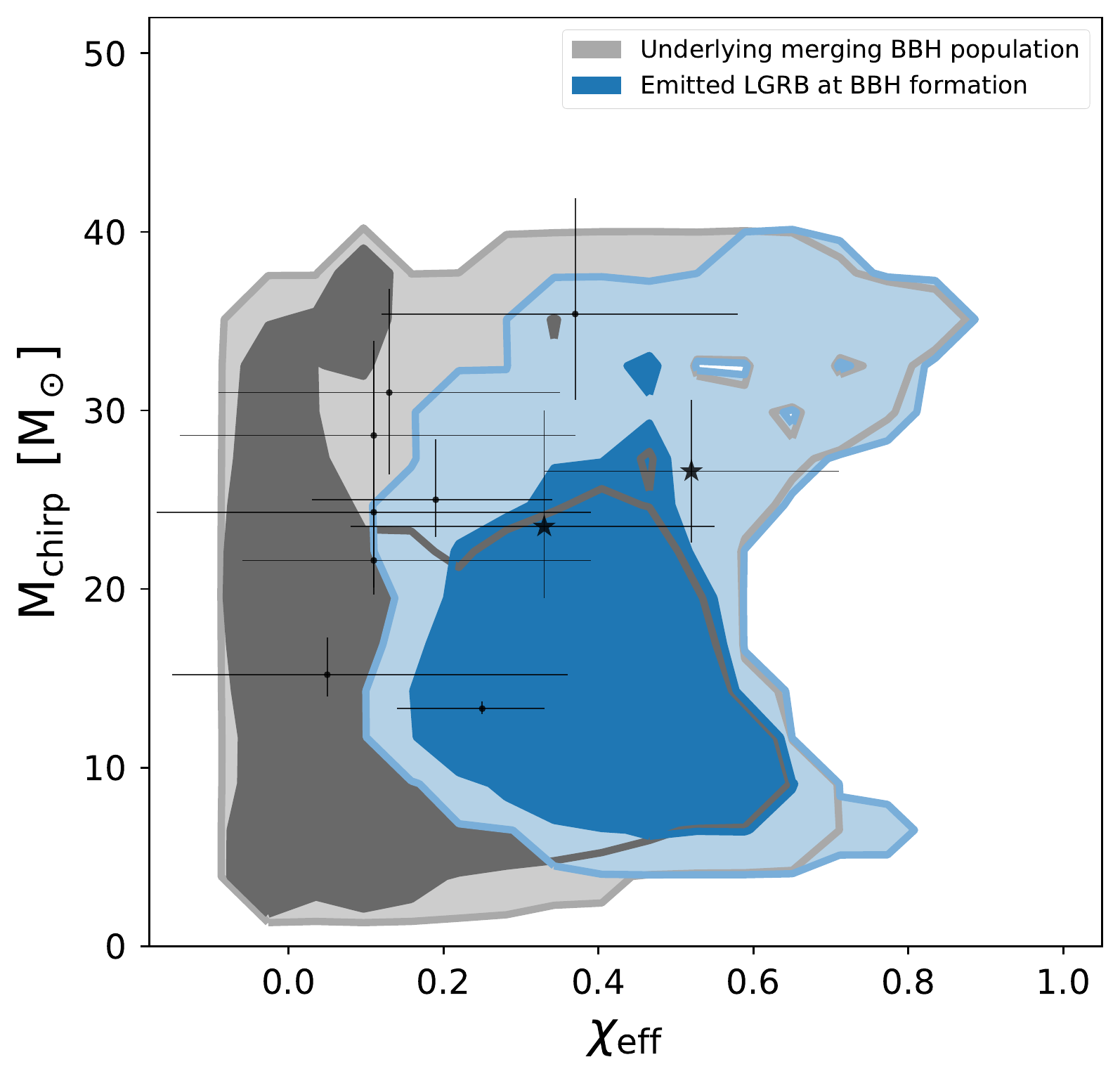}}
\caption{Joint distribution of the chirp mass $\mchirp$ and the effective inspiral spin parameter $\chieff$ for the combined CE, SMT, and CHE channels. 
For all figures, the model predictions for the underlying (intrinsic) BBH population are shown in gray where lighter colors represent larger contour levels of $90\%$ and $99.9\%$, respectively. 
\textit{Left:}  The detected BBH population with O3 sensitivity is shown in orange. Overlaid in black are the O1, O2, and O3a LVC GWTC-2 \citep{2020arXiv201014533T} data with their $90\%$ credible intervals; GW190521 is outside the plotted window.
\textit{Right:} The BBH sub-population which emitted LGRBs at BBH formation is shown in blue. The 10 events in GWTC-2 with chances $>10\%$ to have emitted a luminous LGRB at BBH formation are indicated in black. The 2 events, GW190517 and GW190719, with $>50\%$ probabilities are indicated with star markers. No bin smoothing was applied to construct the contour levels.}
\label{fig:model_vs_GWTC2}
\end{figure*}

Although several formation pathways of coalescing BBHs have been proposed in the literature, 
recent works suggest that the evolution of isolated binaries dominates
the underlying, local merging BBH population \citep{2021ApJ...910..152Z,2021arXiv210503349F,2021arXiv210905836B} over dynamical formation in dense stellar environments \citep[e.g.,][]{2019PhRvD.100d3027R,2019MNRAS.486.5008A} or primordial merging BBHs \citep[e.g.,][]{2016PhRvL.117f1101S, 2020JCAP...06..044D}. However, There is not yet enough observational evidence to make a definite conclusion regarding the origin of BBHs.

The isolated binary formation pathways include (i) a stable mass transfer (MT) and a common envelope (CE) phase \citep[e.g.,][]{
1976ApJ...207..574S,1976IAUS...73...35V,1993MNRAS.260..675T,2007PhR...442...75K,2014LRR....17....3P,2016Natur.534..512B,2020A&A...635A..97B}, (ii) double stable MT (SMT) \citep[e.g.,][]{2017MNRAS.471.4256V,2017MNRAS.468.5020I,2019MNRAS.490.3740N,2021A&A...647A.153B} or (iii) chemically homogeneous evolution (CHE) \citep[e.g.,][]{2009A&A...497..243D,2016MNRAS.458.2634M,2016A&A...588A..50M,2020MNRAS.499.5941D}.  In these channels, high BH spins are the result of tidal spin-up in the BBH progenitor system, which leads to a high AM content in the pre-collapse cores. The high spins of the cores are retained until collapse, even in the case of efficient AM transport \citep{1999A&A...349..189S,2002A&A...381..923S,2019MNRAS.485.3661F}.
In contrast, efficient AM coupling in isolated single-star evolution or in wide binaries is expected to lead to BHs with negligible spin \citep{2018A&A...616A..28Q,2019ApJ...881L...1F} which AM transport efficiency is supported by asteroseismic data \citep{
2014MNRAS.444..102K,2015A&A...580A..96D,2018A&A...616A..24G}, observations of white dwarfs spins \citep{
2005A&A...444..565B} and recent gravitational-wave observations \citep{2021ApJ...910..152Z}.

The collapse of a spinning stellar core has been linked to long-duration gamma-ray bursts (LGRBs) under the ``collapsar'' model \citep{1993ApJ...405..273W,1998ApJ...494L..45P}. In this scenario, portions of the star supported by their extreme AM do not fall directly towards the center when they collapse, forming instead an accretion disk. As the newly-formed central BH accretes from the disk, a fraction of the accreted material's rest mass is converted into energy powering a jet that pierces a hole through the collapsing star's poles, giving rise to the LGRB. Being bright transient events, LGRBs are detectable up to very high redshifts \citep[e.g., $z\approx 9$,][]{2011ApJ...736....7C} and have $T_{90}>2\,\mathrm{s}$, where $T_{90}$ is the time over which a burst emits 90\% of its total measured counts \citep{1993ApJ...413L.101K}. Furthermore, several LGRBs have been associated with Type Ic-broad-line supernovae \citep{ 2006ARA&A..44..507W}. These supernovae show broad spectral lines due to their high kinetic energy and lack H- and He-lines, which indicate that the progenitors are stripped stars \citep{2016ApJ...832..108M}. There are only a few unbiased and redshift-complete catalogs of LGRBs, as they require a rapid follow-up response from the ground to obtain redshift measurements. The largest of these catalogs is the SHOALS survey which counts 110 LGRBs and is considered complete for all LGRBs with fluence $S_{15-150\,\mathrm{keV}} > 10^{-6}\, \mathrm{erg}\,\mathrm{cm}^{-2}$ which corresponds to isotropic-equivalent energies of $E^{\mathrm{iso}}_\mathrm{LGRB} > 10^{51}\,\mathrm{erg}$ in the $45-450\,\mathrm{keV}$ band \citep{2016ApJ...817....7P}.


\begin{figure*}
\centering
\includegraphics[width=0.75\linewidth]{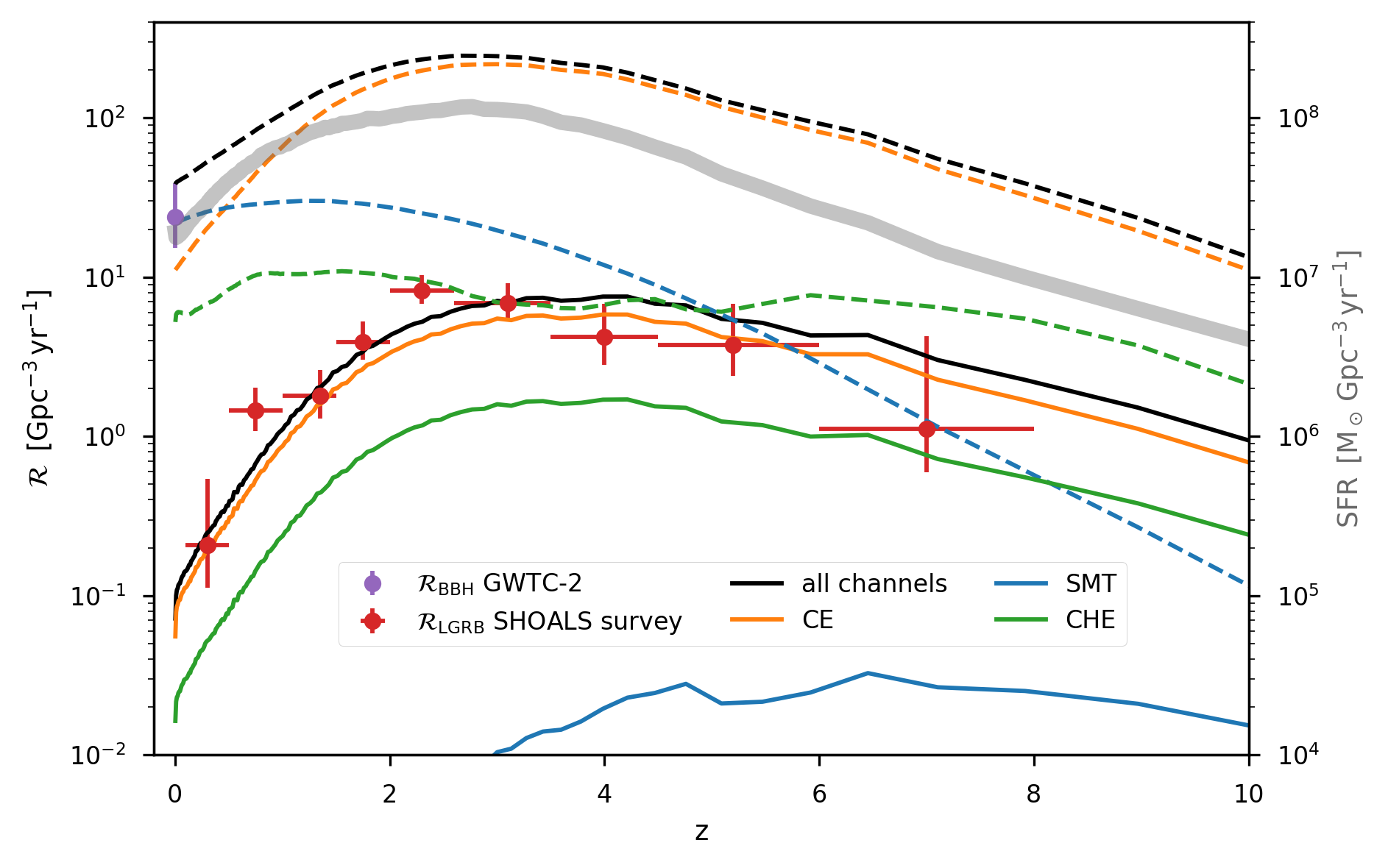} 
\caption{
Modeled merging BBH and luminous LGRB rate densities as a function of redshift from isolated binary evolution in dashed and solid black lines, respectively. The CE, SMT, and CHE channel contributions are indicated in orange, blue, and green colors, respectively. The violet marker denotes observable constraints of local BBH rate densities at $z=0$ from LVC GWTC-2 \citep{2020arXiv201014533T} and the red markers the luminous LGRB rate densities from the SHOALS survey \citep{2016ApJ...817....7P}.
The SHOALS survey LGRB rate densities are not beaming-corrected and hence probe the observed and not the intrinsic LGRB population. Our fiducial model assumes LGRB efficiency $\eta=0.01$, constant beaming factor $f_\mathrm{B}=0.05$, and IllustrisTNG redshift- and metallicity-dependent star formation rate \citep{2019ComAC...6....2N}.
}
\label{fig:R_fiducial}
\end{figure*}

Detailed stellar models of tidally spun-up stars have shown that binary configurations, such as those involved in the formation of fast-spinning merging BBHs from isolated binary scenarios, can lead to LGRBs \citep{2007Ap&SS.311..177V,2008A&A...484..831D,2016A&A...588A..50M,2018A&A...616A..28Q, 2020MNRAS.491.3479C}. Notably, one of the first quantitative studies by \citet{2008A&A...484..831D} concluded that only a small fraction of LGRBs can come from
tidal spin-up, in contrast to findings of more recent studies, including this work.

In this work, we make the working assumption that the isolated binary evolution pathway dominates the formation of merging BBHs in the Universe. We adopt a formation model that combines the CE, SMT, and CHE BBH channels
and is consistent with observed BBH merger rates and their observable distributions \citep{2020MNRAS.499.5941D,2021A&A...647A.153B,2021ApJ...910..152Z}, and explore the hypothesis of a direct link between a potentially significant fraction of the observed long gamma-ray bursts and the progenitors of highly spinning, merging BBHs.

\section{Methods}

The modeling of the BBH population combines detailed binary stellar \texttt{MESA} \citep{2011ApJS..192....3P} models that follow in detail the tidal spin-up of the collapsing cores, with rapid population synthesis techniques \citep{2020ApJ...898...71B} under the same software framework called \texttt{POSYDON}.\footnote{posydon.org} The key assumptions of these models are summarised in Appendicies~\ref{sec:popsyn_models} to \ref{sec:iso-energy-LGRB}. To compute the corresponding rate densities, we assume a redshift- and metallicity-dependent star formation rate (SFR) density according to the IllustrisTNG cosmological simulation \citep{2019ComAC...6....2N} as explained in Appendix~\ref{sec:rates}.

\section{Results}

The combined gravitational-wave (GW) observable predictions of $\chieff$ and $\mchirp$ for the modeled underlying population of merging BBHs is shown in gray in Figure~\ref{fig:model_vs_GWTC2}. 
The CE evolutionary pathway leads to BH--Wolf-Rayet systems in close orbits where a subsequent tidal spin-up phase may occur \citep{2020A&A...635A..97B,2021A&A...647A.153B,2021RNAAS...5..127B}. The SMT channel leads, on average, to wider orbital separations and, hence, the majority of these systems will avoid efficient tidal spin-up \citep{2021A&A...647A.153B}. CHE occurs in initially close binaries with stars that have nearly equal masses and orbital periods between 0.4 and 4 days \citep{2020MNRAS.499.5941D}. Both stars experience strong tidal spin-up since early in their evolution, which leads to efficient rotational mixing throughout their interior, avoiding a super-giant phase and associated stellar expansion. Therefore, the CE and CHE scenarios are mostly responsible for BBHs with non-zero $\chieff$ \citep{2020A&A...635A..97B,2021A&A...647A.153B}, where the CHE BBHs primarily probe high $\mchirp$ \citep{2020MNRAS.499.5941D}.

Contemporary GW detectors can probe only the low redshift subset  \citep[$z\lesssim 1$,][]{2020arXiv201014527A} of the underlying BBH population. Observations are biased towards high $\mchirp$ as the signals of massive BBHs are louder and, hence, can be detected at further distances. Current GW observatories are therefore unable to individually resolve a large fraction of merging BBHs in the Universe. In the left panel of Figure~\ref{fig:model_vs_GWTC2}, we indicate in orange the observed distribution of $\chieff$ and $\mchirp$ predicted by our model, assuming a three detector configuration with a network signal-to-noise ratio threshold of $12$ and ``mid-high/late-low'' sensitivity \citep{2018LRR....21....3A}, consistent with the third observing run of LIGO and Virgo detectors. For a direct comparison with the observations, we overlay the 46 BBH events with their 90\% credible interval (CI) in black. The GW detector selection effects distort the observable distributions to high $\mchirp$ and $\chieff$ values compared to the underlying BBH distribution, which is shown in gray.

A fraction of the underlying merging BBH population with highly spinning BHs is expected to give rise to LGRB events at the moment of BBH formation. For each BBH formation, we calculate from the structure profile of the BH progenitor star whether a sufficiently massive accretion disk is formed during the core collapse, which will give rise to a luminous LGRB (see Appendix~\ref{sec:iso-energy-LGRB} for details).  In the CE channel, only the second-born BH is associated with a LGRB as tidal interactions are only relevant in the BH--Wolf-Rayet evolution phase of the BBH progenitor. In contrast, a highly spinning CHE BBH system can be associated with two LGRB events, as tides cause both stars to be rapidly spinning. The sub-population of BBHs associated with LGRBs is indicated in blue in the right panel of Figure~\ref{fig:model_vs_GWTC2}. These systems have $\chieff\gtrsim 0.2$ (90\% CI) while favoring $\mchirp \in [5,30]\,\Msun$. In contrast to the observed GW population, there is no observational bias for high-$\mchirp$ BHs in the LGRB population. We find that the expected number of GWTC-2 events that had emitted a LGRB at BBH formation is $\approx 4$. Among all the GWTC-2 events, GW190517 and GW190719 have the highest probabilities, $\approx$85\% and $\approx$60\% respectively, of having had a LGRB precursor, while 8 more events have a probability $p_\mathrm{LGRB}>10\%$. Those 10 events are highlighted in the right panel of Figure~\ref{fig:model_vs_GWTC2}. The details of the calculation of these probabilities are presented in Appendix~\ref{sec:LGRB_prob}.


The combined local ($z=0$) BBH merger rate density of our CE, SMT, and CHE fiducial models is $38.3\,\rdunits$, with each channel contributing 57\%, 29\%, and 14\%, respectively.
The predicted local rate density is within the observational constraints from GWTC-2 with $[15.3,38.8]\,\rdunits$ at 90\% credibility \citep{2020arXiv201014533T}.
In Figure~\ref{fig:R_fiducial}, we show the redshift evolution of each channel's BBH merger rate density as well as their combination (dashed lines). The CE BBH merger rate density peaks at a redshift $z \in [2,3]$, close to the peak of the SFR density, shown in gray. The CE BBH merger rate closely follows the SFR because of the short delay times between the formation and merger of tight BBH systems produced by the CE channel. In contrast, SMT and CHE BBHs have longer delay timescales as there is no mechanism to shrink the orbits as efficiently as the CE phase does. Therefore, the SMT rate density does not follow the SFR density and peaks at lower redshifts. Finally, we note that the CHE rate density is not as suppressed at high redshift as in the other two channels. This is because the CHE channel operates with higher efficiency at extremely low metallicity environments, which are more abundant at high redshifts.

\begin{figure}[ht]
\centering
\subfloat{\includegraphics[width=\linewidth]{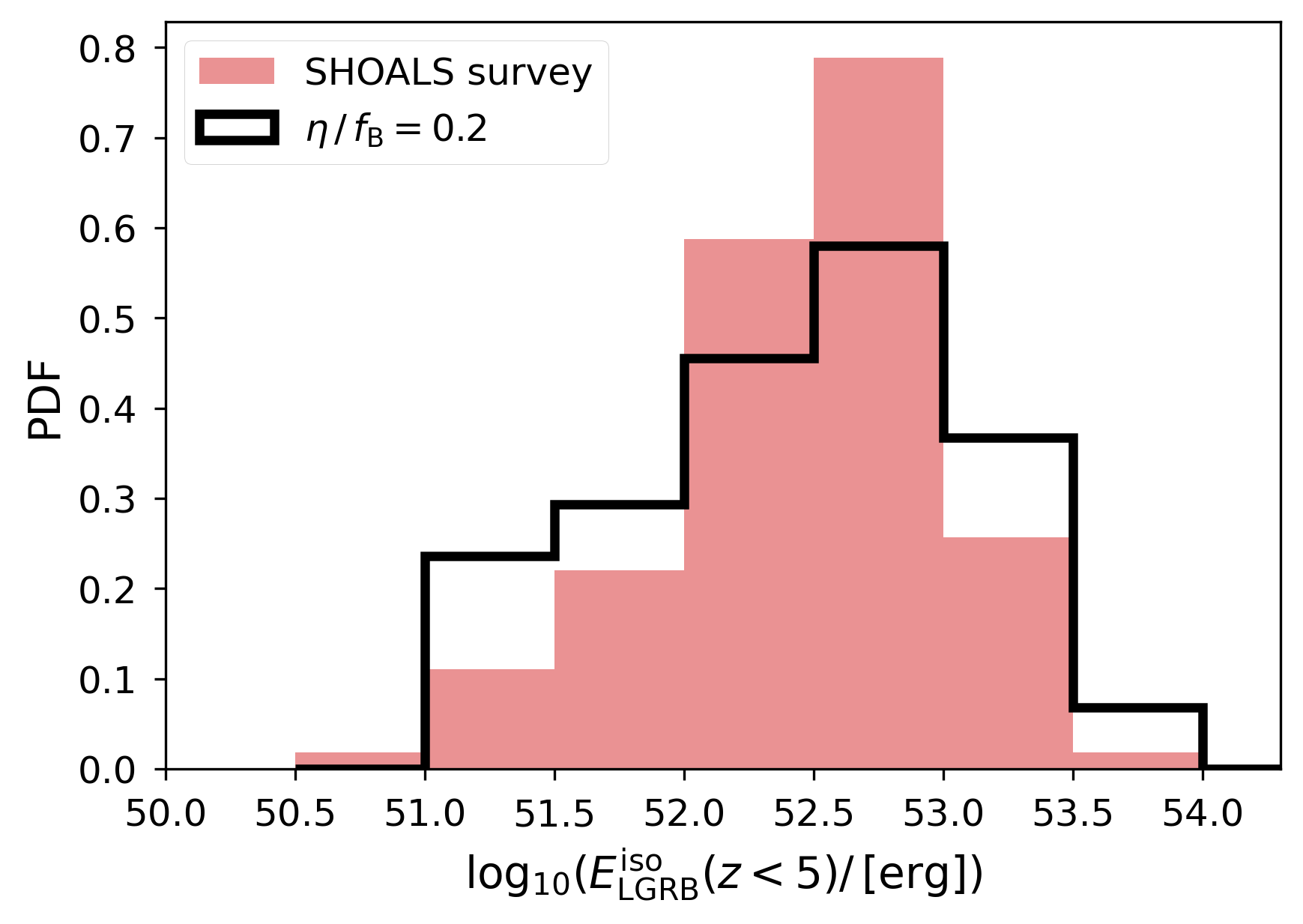}}
\caption{Normalized histogram of the observed luminous LGRB isotropic-equivalent energies with redshift $z<5$ from the SHOALS survey, in light red, compared to the modeled LGRB isotropic-equivalent energies. Our fiducial model was calibrated such that the modeled LGRB energies peak near the observed energy distribution. This is achieved for  $\eta/f_\mathrm{B} = 0.2 \propto E_\mathrm{LGRB}^\mathrm{iso}$. 
} 
\label{fig:energies}
\end{figure}

Luminous LGRB rate densities from our fiducial model are shown in Figure~\ref{fig:R_fiducial} as a function of redshift, for each channel and their combination (solid lines). The fiducial model assumes a LGRB energy efficiency $\eta=0.01$ and beaming fraction $f_\mathrm{B}=0.05$, whose ratio is calibrated to match the peak of observed luminous LGRB energy distributions as described in Appendix~\ref{sec:iso-energy-LGRB} and shown in Figure~\ref{fig:energies}. The majority of LGRBs originate through the CE evolutionary pathway while only 21-25\%, for any $z<10$, come from CHE. The SMT channel leads to the smallest LGRB rate densities ($<0.03 \,\mathrm{Gpc}^{-3}\,\mathrm{yr}^{-1}$) for any redshift, as tidally spun-up second-born BHs are rare in this evolutionary pathway. 
To confront our model predictions, we compare our theoretical luminous LGRB rate estimates with the SHOALS survey estimates using red markers in  Figure~\ref{fig:R_fiducial}. The combination of CE and CHE LGRB rates for our fiducial model are consistent with the observations of luminous LGRBs throughout the redshift range. A discussion about the sensitivity of our rate estimates about the choice of beaming fraction and SFR are presented in Appendices~\ref{sec:iso-energy-LGRB} and \ref{sec:rates}.

\begin{figure}
\centering
\includegraphics[width=\linewidth]{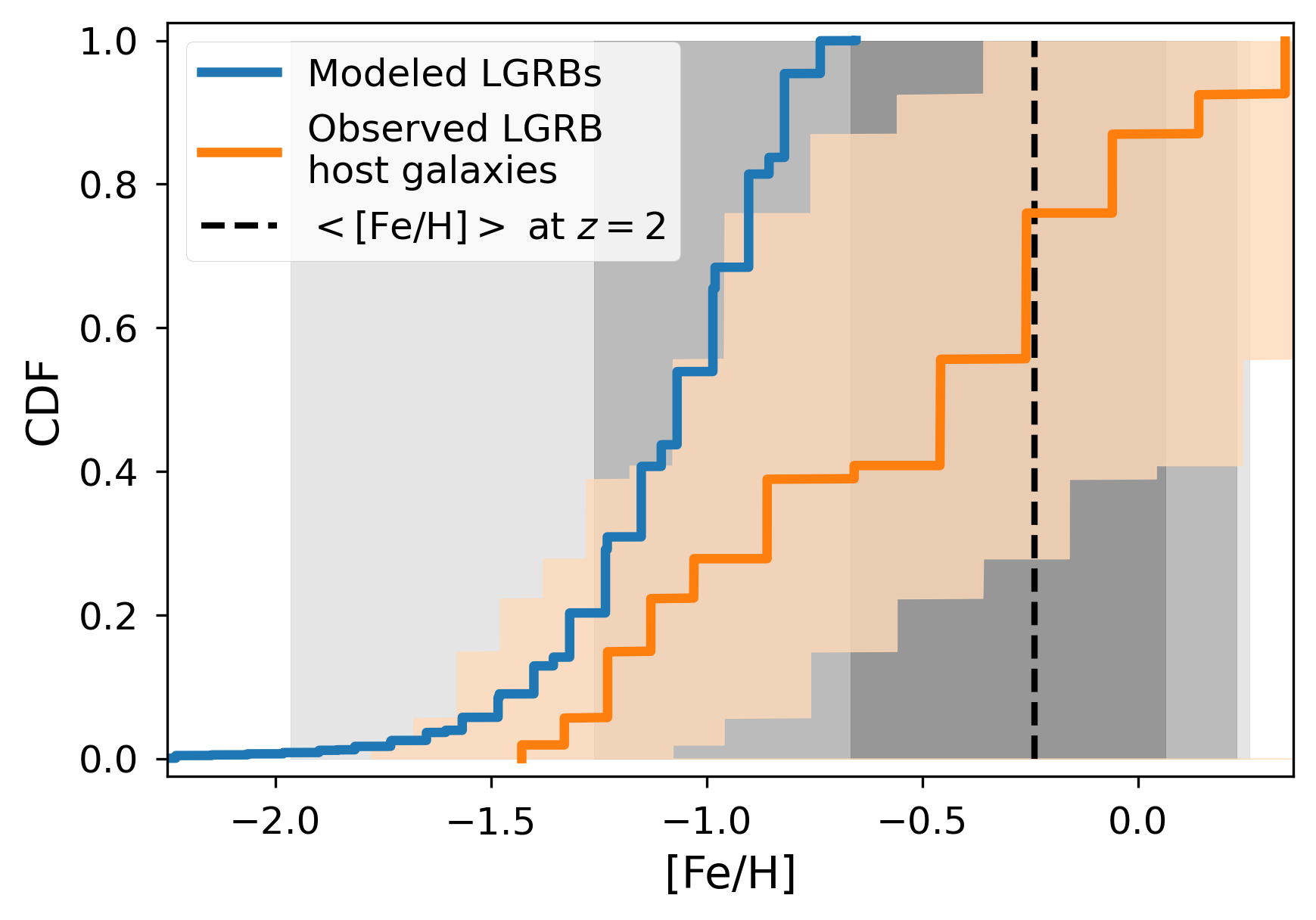} 
\caption{
Cumulative distribution function (CDF) of the modeled LGRB progenitors' metallicities for redshifts $z<2.5$, in blue. The CDF of the observed SHOALS LGRBs host galaxy metallicities for $z<2.5$ \citep{2019arXiv190402673G} are indicated in orange. The light orange shaded area shows the uncertainty in the observed CDF due to  systematic offsets in the measurement of $\log_{10}(\mathrm{O/H})$ depending on the calibrations used, and the stellar mass of the galaxy which can be as high as $\Delta[\log_{10}(\mathrm{O/H})] \approx 0.7\,\mathrm{dex}$ \citep{2008ApJ...681.1183K}. As a reference, we indicate with a vertical dashed black line the median metallicity from the IllustrisTNG simulation at redshift $z=2$ and lighter gray shaded areas delineate larger CI of 68, 95 and 99\% for the assumed star formation metallicity distribution.
}
\label{fig:metallicity}
\end{figure}

LGRBs probe the formation of highly spinning merging BBHs formed at low metallicity because, at such metallicities, stellar winds are weaker, which allows the BBHs' progenitors to remain rapidly spinning and in close orbits until the formation of the BHs. These systems are therefore mostly formed at high redshifts where low metallicity environments are more abundant. Measurements of the metallicity of LGRB host galaxies have shown that LGRB rates are indeed enhanced at low metallicities \citep{2006Natur.441..463F}. In our model, the threshold for LGRB formation is $Z_\mathrm{max}\approx0.2\,Z_\odot$ where we adopt $Z_\odot=0.017$ \citep{1996ASPC...99..117G}. In Figure~\ref{fig:metallicity}, we compare the progenitors' metallicities of modeled LGRBs to the sub-sample of the SHOALS LGRBs with identified host galaxies which have measured metallicities for $z<2.5$ \citep{2019arXiv190402673G}. At face value, we find that 40\% of the observed LGRB host galaxies have metallicities lower than $Z_\mathrm{max}$. However, when taking into account possible systematic uncertainties in the measurement of $\log_{10}(\mathrm{O/H})$ abundances \citep{2008ApJ...681.1183K} our model can be consistent with at most $\sim$85\% of observed LGRBs, see Appendix~\ref{sec:metallicity} for more details. Selection effects in LGRB host galaxies for which metallicity measurements are possible, biases the sample towards low red-shift and high-mass galaxies, and hence potentially towards higher metallicities \citep{2019arXiv190402673G}. This comparison implies that in order to associate the entirety of luminous LGRBs to the formation of BBHs, a potentially significant fraction of LGRB progenitors at low redshifts ($z<2.5$) must originate in low metallicity pockets of the host galaxies. Finally, we should stress that theoretical model uncertainties due to the uncertain metallicity dependence of stellar wind mass loss during the late Wolf-Rayet phase of the stars evolution as well as the uncertainties in the metallicity dependent SFR make this comparison less conclusive. A detailed parameter study would improve such a comparison but is outside the scope of this letter.

\section{Discussion \& Conclusions}

In this study, we only consider a contribution to the LGRB rate from merging BBH progenitors. Other pathways to fast-spinning, BH progenitor stars, in single or binary stars, have been proposed to lead to LGRBs \citep[e.g.,][]{2006A&A...460..199Y,2007A&A...465L..29C}, none of which though at a rate that matches the observed one, when considering efficient angular momentum transport in stellar interiors \citep{2007PASP..119.1211F}. Observed high mass X-ray binaries, containing highly spinning BHs, such as Cygnus~X-1 \citep[e.g.,][]{2011ApJ...742...85G,2021ApJ...908..117Z}, may have also been progenitors of LGRBs. The formation of these systems is puzzling \citep[e.g.][]{2012ApJ...747..111W,2021ApJ...908..118N}, and it is  uncertain whether the BH spin in these systems originates from highly-spinning pre-collapse cores \citep[see e.g.,][]{2008ApJ...689L...9M,2017ApJ...846L..15B,2018ApJ...862L...3S}. It is interesting to note that a simple estimate of the LGRB rate density from Cyg~X-1 like systems, assuming that there is one such binary per Milky Way like galaxy with $\mathrm{SFR}_\mathrm{MW} \simeq 1\,\mathrm{M}_\odot\,\mathrm{yr}^{-1}$, and a typical lifetime of $\tau_\mathrm{HMXB} \simeq 10^5\,\mathrm{yr}$, far exceeds the observationally determined one at $\mathcal{R}_\mathrm{LGRB}(z\simeq 0) < 0.6 \, \mathrm{Gpc}^{-3} \,\mathrm{yr}^{-1}$ at
\begin{equation}
\mathcal{R}_\mathrm{LGRB}^\mathrm{HMXB}(z \simeq 0) = 10\times\left(\frac{\tau_\mathrm{HMXB}}{10^5\,\rm yr}\right)\times\left(\frac{f_\mathrm{B}}{0.05}\right)\,\mathrm{Gpc}^{-3}\,\mathrm{yr}^{-1} \, .
\end{equation}
In this estimate, we assumed a $\mathrm{SFR}(z\simeq 0) = 2 \times 10^{7} \, \mathrm{M}_\odot \, \mathrm{Gpc}^{-3} \, \mathrm{yr}^{-1}$ for the local Universe (cf. Figure~\ref{fig:model_vs_GWTC2}). Another possible viable alternative for the origin of LGRBs includes the formation of a fast rotating neutron star with an ultrahigh magnetic field \citep{1992ApJ...392L...9D}.
While our analysis cannot exclude other potential progenitors of LGRBs, consideration of the salient uncertainties of our model demonstrates that progenitors of fast-spinning BBH mergers, formed via isolated binary evolution, are likely a major contribution to the observed luminous LGRB rate.

Fast-spinning BBHs have typically short merger timescales. Because of this, current gravitational wave detectors cannot probe them efficiently, as their formation and merger rate is maximal approximately where the SFR density peaks at $z \in [2,3]$. Luminous LGRBs, on the other hand, are observable  up to redshift of $\approx9$, and can therefore be used as a cosmological probe, empirically constraining the sub-population of progenitors of fast-spinning BBH merger events far beyond the horizons of current-generation gravitational wave observatories. We have used two types of multi-messenger, albeit asynchronous, types of observations, gravitational waves, and gamma-rays, to chart BBH formation across cosmic time. Using combinations of observations like this opens a new avenue to constrain the currently uncertain physics of binary evolution and compact object formation.

\textbf{Acknowledgements } We would like to thank Christopher Berry for his comments on the manuscript. This work was supported by the Swiss National Science Foundation Professorship grant (project number PP00P2\_176868).
E.R.-R. acknowledges support from  the Heising-Simons Foundation, the Danish National Research Foundation (DNRF132) and NSF (AST-1911206 and AST-1852393). P.M. is supported by the FWO junior postdoctoral fellowship No.\ 12ZY520N. L.Z.K. is supported by CIERA. M.Z. is supported by NASA through the NASA Hubble Fellowship grant HST-HF2-51474.001-A awarded by the Space Telescope Science Institute, which is operated by the Association of Universities for Research in Astronomy, Inc., for NASA, under contract NAS5-26555. J.J.A. and S.C. are supported by CIERA and A.D., J.G.S.P., and K.A.R. are supported by the Gordon and Betty Moore Foundation through grant GBMF8477. Y.Q. acknowledges funding from the Swiss National Science Foundation under grant P2GEP2{\_}188242. The computations were performed in part at the University of Geneva on the Baobab and Yggdrasil computer clusters and at Northwestern University on the Trident computer cluster (the latter funded by the GBMF8477 grant). All figures were made with the open-source Python modules \texttt{Corner} \citep{corner} and \texttt{Matplotlib} \citep{Hunter:2007}. This research made use of the python modules 
\citep{2020SciPy-NMeth}, \texttt{Astropy} \citep{price2018astropy}, and \texttt{PyCBC} \citep{alex_nitz_2019_3546372}.


\bibliography{aanda}

\appendix

\onecolumn

\section{Population synthesis of CE, SMT, and CHE binary black holes}\label{sec:popsyn_models}

We model the evolution of binaries through CE and SMT with the \texttt{POSYDON} framework to combine the rapid population synthesis code \texttt{COSMIC} \citep{2020ApJ...898...71B} with detailed $\texttt{MESA}$ \citep{2011ApJS..192....3P,2013ApJS..208....4P, 2015ApJS..220...15P,2018ApJS..234...34P,2019ApJS..243...10P} stellar structure and binary evolution simulations \citep{2021A&A...647A.153B}. This hybrid approach is used to rapidly evolve millions of binaries from the zero-age main sequence (ZAMS) until the end of the second MT episode. For the last phase of the evolution, which determines the second-born BH spin \citep{2018A&A...616A..28Q,2020A&A...635A..97B}, we used detailed BH--Wolf-Rayet binary evolution simulations to model the tidal spin-up phase until the secondary star reached central carbon exhaustion. These simulations take into account differential stellar rotation, tides, stellar winds, and the evolution of the Wolf–Rayet stellar structure until carbon depletion. The core collapse is modeled as described in the next section. We consider disk formation during the collapse of highly spinning stars, mass loss through neutrinos, pulsational pair-instability and pair-instability supernova (PPISN \& PISN) \citep{2019ApJ...882...36M}, and orbital changes resulting from anisotropic mass loss and isotropic neutrinos mass loss \citep{1996ApJ...471..352K}.

In our models, the first-born BHs in the SMT and CE channels are formed with a negligible spin because of the assumed efficient AM transport \citep{2015ApJ...800...17F,2018A&A...616A..28Q,2019ApJ...881L...1F}. If AM transport were to be inefficient, this would lead to spinning BBHs \citep{2020A&A...636A.104B}, which are currently inconsistent with GWTC-2 observations. Moreover, we assume Eddington-limited accretion efficiency onto compact objects, resulting in a negligible amount of mass accreted onto the first-born BH during SMT. Hence, the first-born BH in the SMT channel avoids any spin-up during MT \citep{1974ApJ...191..507T}. Alternatively, if the accretion onto compact objects could reach highly super-Eddington rates, the binaries would not shrink enough to produce merging BBHs, leading to the suppression of the SMT channel \citep{2021A&A...647A.153B}. Hence, even though super-Eddington accretion efficiency strongly affects the yield of merging BBHs through the SMT channel, it will not affect LGRBs rates as the MT accretion spin-up occurs after BH formation. Finally, motivated by the model comparison between our models and GWTC-2 data \citep{2021A&A...647A.153B}, we assume inefficient common envelope ejection efficiencies, taken as $\alpha_\mathrm{CE}=0.5$ in the $\alpha_\mathrm{CE}-\lambda$ CE parameterization theory \citep[see, e.g.][for a review]{2013A&ARv..21...59I} and adopt $\lambda$ fits as in \citet{2014A&A...563A..83C}. Because the orbital separation post CE is approximately proportional to $\alpha_\mathrm{CE}$, inefficient CE ejection leads, on average, to a larger fraction of tidally spun-up BHs, but at the same time to a smaller overall number of BBH merger events compared to efficient CE ejection, $\alpha_\mathrm{CE} > 1$. Here, an $\alpha_\mathrm{CE}$ value grater than 1 does not mean that other sources of energy partake in the CE ejection, but more likely points to an inaccurate assumption of core-envelope boundaries. Indeed, multiple recent studies \citep{2019ApJ...883L..45F,2019A&A...628A..19Q,2021A&A...645A..54K,2021A&A...650A.107M} have shown that envelope stripping stops earlier than what currently assumed in population synthesis. We find that this model's uncertainty changes our LGRB rate estimate by $\mathcal{R}_\mathrm{LGRB}^{\alpha_\mathrm{CE}=0.5}$ at redshift $z=0$ ($z=2$) by $+36$\% ($+18$\%), $-56$\% ($-42$\%) and $-68$\% ($-54$\%) for $\alpha_\mathrm{CE}=0.25,1$ and $2$, respectively, not changing our study's conclusion.

The binary evolution through CHE is modeled entirely with \texttt{MESA} until the carbon depletion of both stars \citep{2020MNRAS.499.5941D}. More precisely, we model the two stars simultaneously in a binary system where tidal interaction and mass transfer are taken into account. For consistency, the CE and SMT \texttt{MESA} models used identical input physics to the CHE ones, while simulations with the \texttt{COSMIC} code were also configured to be as consistent as possible \citep{2021A&A...647A.153B,2021ApJ...910..152Z}. Similar to the other channels, the stars' profiles' core collapse is done self-consistently with CE and SMT models using \texttt{POSYDON}. Because the CHE \texttt{MESA} grids assume a fixed mass ratio $q=1$, both stars will reach core collapse simultaneously. In practice, we collapse one star after the other by applying a Blauw kick \citep{1996ApJ...471..352K} after each star has collapsed to account for the orbit adjustment resulting from PPISN and neutrinos mass loss, where we assume circularization after the formation of the first BH \citep{2020MNRAS.499.5941D}.

Initial binary conditions at ZAMS are drawn randomly from empirically constrained distributions. In CE and SMT, the ZAMS binaries are directly evolved with \texttt{POSYDON} while binaries in the parameter space leading to CHE are mapped to the nearest neighbor CHE \texttt{MESA} evolutionary track. Metallicities are sampled in the log-range $\log_{10}(Z)\in[-5,\log_{10}(2Z_\odot)]$. For the CE and SMT models the log-metallicity range is divided in 30 desecrate values from $\log_{10}(Z) =-4$ to $\log_{10}(1.5Z_\odot)$ where binaries with $\log_{10}(Z)\in[-5,-4]$ are mapped to the lowest metallicity bin \citep{2021A&A...647A.153B}. For the CHE models the log-metallicity range is sampled with 22 discrete values from $\log_{10}(Z) = -5.0$ to $\log_{10}(Z) = -2.375$, above which any binary evolves through the CHE channel \citep{2020MNRAS.499.5941D}. Primary masses follow the Kroupa initial mass function (IMF), a broken power law with coefficient $\alpha=-2.3$ \citep{2001MNRAS.322..231K} in the sampled mass range
$5\,\mathrm{M}_\odot \leq m_1 \leq 150 \,\mathrm{M}_\odot$. The upper limit is an extrapolation of the original Kroupa IMF measured only up to $50\,\mathrm{M}_\odot$. The arbitrary maximum stellar mass is chosen to exclude BH formation above the upper mass gap of PISN, which we do not model \citep{2002luml.conf..369H}. The mass distribution of the less massive secondary star is given by $m_2 = m_1 \times q$, where the initial mass ratio $q$ is drawn from a flat distribution \citep{2012Sci...337..444S} in the range $q \in (0, 1]$. We assume that all binaries are born with circular
orbits. Furthermore, we adopt a binary fraction of $f_\mathrm{bin} = 0.7$ \citep{2012Sci...337..444S} and assume that at birth the distribution of log-orbital periods follow a power law with coefficient $\pi = -0.55$ \citep{2012Sci...337..444S} in the range $\log_{10}(p/[\mathrm{day}]) \in [0.15, 5.5]$ and extrapolate down to the range $\log_{10}(p/[\mathrm{day}]) \in [\log_{10}(0.4/[\mathrm{day}]), 0.15]$ assuming a log-flat distribution \citep{2021A&A...647A.153B}. The portion of the parameter space with $q\in[0.8,1]$ and $p \in [0.4, 4]$ days may lead to CHE \citep{2020MNRAS.499.5941D}. Notice that there are some uncertainties on the actual initial binary properties of mass ratios, periods, and eccentricities \citep[see e.g.][]{2017ApJS..230...15M}, however, there are no constraints on them at low metallicities such as the one modeled here. Moreover, The extrapolation to low orbital periods causes us to sample systems Roche-lobe overflowing at ZAMS. Therefore, these systems have undergone MT during the pre-main sequence phase, which complicates the binary evolution and, a priori, might not lead to CHE. To remove these systems from the sampled distribution, we adopt ZAMS stellar radii fits \citep{1996MNRAS.281..257T}, which we compare to the initial Roche-lobe radii of the binary \citep{1983ApJ...268..368E}. The population synthesis will then result in a synthetic population of BBHs, which we distribute across the cosmic history of the Universe to compute rate densities. See the later section for a detailed description.

\section{LGRB collapsar scenario}\label{sec:collapsar}

A massive star collapses under its own weight when nuclear reactions can no longer generate enough pressure to balance the pull of gravity. For the most massive stars, this occurs after the stars have formed iron cores. Due to computational constraints, our \texttt{MESA} simulations run until carbon depletion, which occurs less than a year before the actual core collapse. Because the remaining stellar evolutionary phase is so rapid compared to the star's total evolution, we can assume that the star's structure will not change drastically in the neglected portion of the evolution. The core collapse is modeled using fits to the results of 2D core-collapse models \citep{2012ApJ...749...91F}. We also account for mass loss through PPISN or stellar disruption from PISN using fits to 1D stellar models targeting this evolution phase \citep{2019ApJ...882...36M}. Depending on the carbon-oxygen core mass, $m_\mathrm{CO-core}$, the star might explode as a supernova and have a fraction of the ejected mass falling back onto the compact object or, if the star is massive enough, where $m_\mathrm{CO-core} \geq 11\,\mathrm{M}_\odot$, the star will collapse directly to form a BH \citep{2012ApJ...749...91F}. Consequently, in our models, only stars with $m_\mathrm{CO-core} \leq 11\,\mathrm{M}_\odot$ can receive natal kicks, with magnitudes drawn from a Maxwellian distribution with $\sigma=265$ km/s \citep{2005MNRAS.360..974H} and rescaled by one minus the fall-back mass fraction \citep{2012ApJ...749...91F}. In this case, only a negligible fraction of such low mass merging highly spinning BBHs associated to LGRBs will be disrupted by natal kicks as they are in tight orbits (orbital periods of less than one day) and, hence, only a kick with magnitude larger than the corresponding orbital velocity $v_\mathrm{orb} > 500\,\mathrm{km/s}$ can disrupt the system. Furthermore, notice that newer studies on core-collapse physics \citep[e.g.][]{2015ApJ...801...90P,2016ApJ...821...38S,2020MNRAS.499.2803P,2021A&A...645A...5S} indicate that there is no such distinct monotonic relation between neutron-star (NS) and BH formation (for a detailed study of the impact of newer core-collapse mechanism prescriptions on the formation of merging BBH and BH-NS in our models, see \citet{2021ApJ...912L..23R}). In the collapse, we also account for up to $0.5\,\mathrm{M}_\odot$ mass loss through neutrinos \citep{2020ApJ...899L...1Z}. If the collapsing star is rapidly rotating, an accretion disk might form during this process \citep{2021A&A...647A.153B}. Because our \texttt{MESA} simulations provide us with the star's profile at core collapse, we can estimate the amount of material that forms an accretion disk around the newly-formed BH and the spin of the final BH \citep{2019arXiv190404835B}. We assume that the innermost shells of the star form a central BH of mass $2.5\,\mathrm{M}_\odot$ through direct collapse, where we account for the mass and AM loss through neutrinos \citep{2021A&A...647A.153B}. The collapse of each subsequent shell happens on a dynamical timescale. We account for each shell's portion with enough specific AM to support disk formation instead of collapsing directly. The thin disk is subsequently accreted on a viscous timescale which we assume to be much smaller than the dynamical timescale. Hence the disk is accreted before the next shell collapses. Notice that the accretion problem might be more complex than what assumed, e.g. \citet{2011MNRAS.410.2385T} 3D smoothed-particle hydrodynamics simulations showed that hydrodynamical instabilities in the accretion disk may result in intermittent accretion. If this is the case one would also need to account for feedbacks from the already-accreted disk to the rest of the in-falling material \citep[see e.g.][]{2021RNAAS...5..127B} which we do not account here. When an accretion disk is formed, a fraction of its rest-mass energy can power the formation of a jet that pierces through the star and breaks out from its poles. This mechanism is known as the collapsar scenario and is thought to give rise to LGRBs \citep{1993ApJ...405..273W,1998ApJ...494L..45P}.

\section{LGRB isotropic-equivalent energy calibration}\label{sec:iso-energy-LGRB}

\begin{figure}[ht]
\centering
\subfloat{\includegraphics[width=0.7\linewidth]{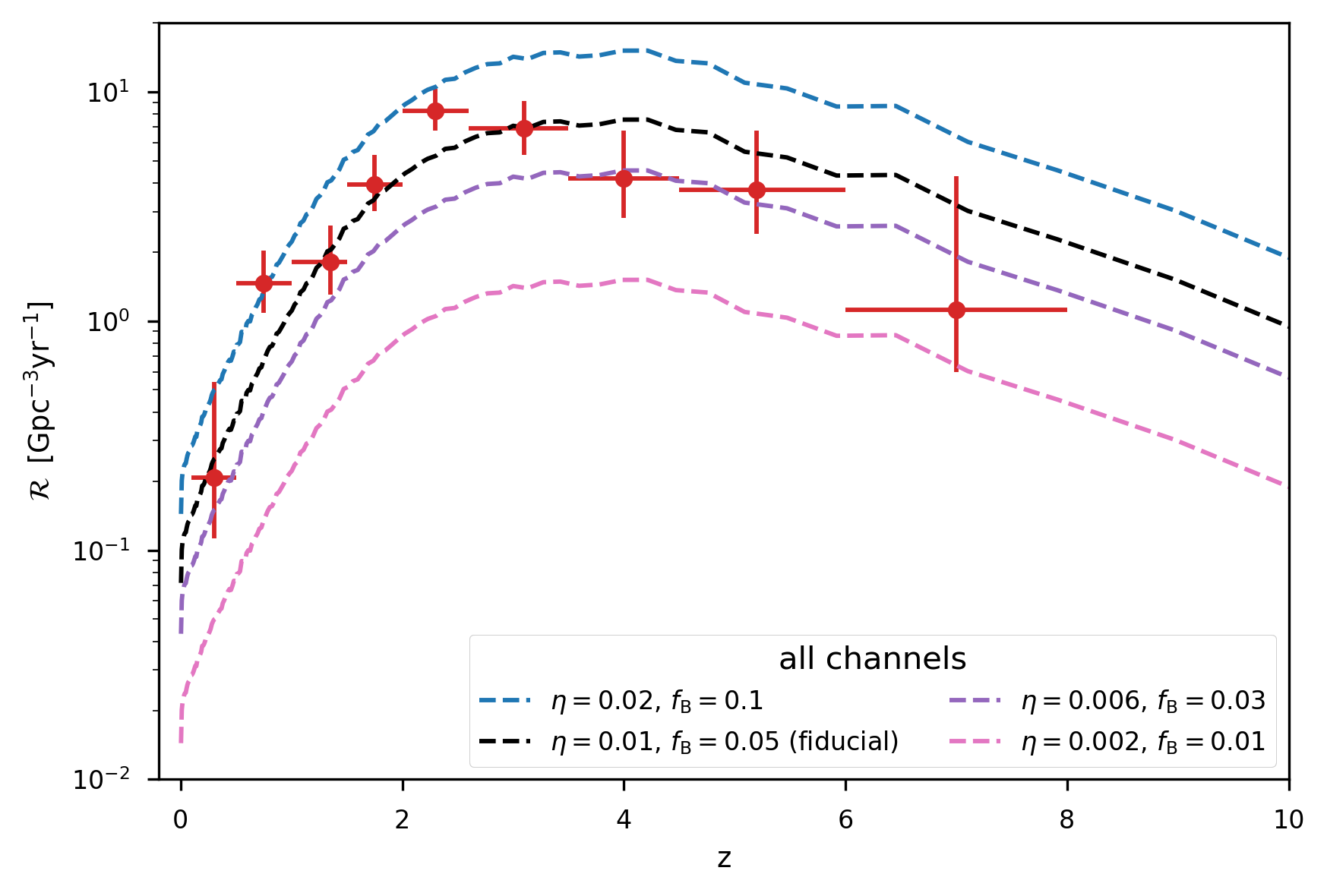}}
\caption{Modeled luminous LGRB rate densities as a function of redshift for all channels combined. The figure illustrates model uncertainties given an arbitrary choice of beaming fraction $f_\mathrm{B}\in[0.01,0.03,0.05,0.1]$. The LGRB energy efficiency $\eta$ is obtained from the isotropic-equivalent energy calibration condition $\eta/f_\mathrm{B}=0.2$.
}
\label{fig:rate_uncertainties}
\end{figure}

The LGRB jet is powered by the accretion disk produced in the core-collapse, and only a fraction, $f_\mathrm{jet}$, of this rest-mass energy will power the jet, of which a fraction $f_\gamma$ is observed in the $\gamma$-ray band $45-450\,\mathrm{keV}$. Moreover, when the jet breaks out from the poles, the star's outer layers, which have yet to collapse, could become unbound by the shock caused by the jet, 
using a fraction of the estimated energy to unbind the star while the rest escapes. Similarly, we can encompass this uncertainty in the parameter $1-f_\mathrm{unbound}$. For simplicity, in our models, we parameterize our ignorance about these processes in the fixed efficiency parameter $\eta=f_\mathrm{jet} \times f_\gamma \times (1-f_\mathrm{unbound})$. 
Hence, the total LGRB energy released in the $\gamma$-ray band by the BH formation process is then
\begin{equation}
    E_\mathrm{LGRB} = \eta \Delta M_\mathrm{disk} \, c^2 \, \, \,  \mathrm{ergs} ,
\end{equation}
where $\Delta M_\mathrm{disk}=\sum_i (1-[1-2GM_\mathrm{BH}/(3c^2r_\mathrm{ISCO,i})]^{1/2})m_\mathrm{disk,i}$ is the total rest mass released as energy during the accretion process which depends on the radius of the innermost stable circular orbit (ISCO) of the accreting central BH, $r_\mathrm{ISCO}$ \citep{1970Natur.226...64B,1974ApJ...191..507T}. Here, $m_\mathrm{disk,i}=m_\mathrm{shell,i} \cos(\theta_\mathrm{disk,i})$ is the mass of the disk formed during the collapse of the $i$th shell with radius $r$ where $\theta_\mathrm{disk,i}$ is the polar angle above which disk formation occurs. This quantity depends on the specific AM of the ISCO of the accreting BH, $j_\mathrm{ISCO}$, and the shell's specific AM, $\Omega(r)r^2$, as
\begin{equation}
    \theta_\mathrm{disk,i} \equiv  \theta_\mathrm{disk}(r) = \arcsin{\left[\left( \frac{j_\mathrm{ISCO}}{\Omega(r) r^2} \right)^{1/2}\right] }\, .
\end{equation}
The jet escapes from the poles and is beamed with a half-opening angle $\theta_\mathrm{B}$. The chance of having the line of sight aligned with the jets is then $f_\mathrm{B} = 1 - \cos(\theta_\mathrm{B})$. 
The total isotropic-equivalent energy released by the LGRB jet is
    \begin{equation}
    E_\mathrm{LGRB}^\mathrm{iso} = f_\mathrm{B}^{-1}  E_\mathrm{LGRB} = f_\mathrm{B}^{-1} \, \eta \, \Delta M_\mathrm{disk,\, rad} \, c^2 \, \, \,  \mathrm{erg} \, .
    \label{eq:ELGRBiso}
\end{equation}
We have two apparent free parameters, $f_\mathrm{B}$ and $\eta$, to determine. For simplicity, we assume that both parameters are constants. We can then use observations of luminous LGRBs from the SHOALS survey to calibrate the ratio $\eta/f_\mathrm{B} \propto E_\mathrm{LGRB}^\mathrm{iso}$ such that the peak of the modeled isotropic-equivalent energy distribution matches the observed one. In Figure~\ref{fig:energies} we show the result of this calibration, namely $\eta/f_\mathrm{B}=0.2$. With this constraint, we can choose reasonable values of $f_\mathrm{B}$ and obtain a corresponding $\eta$. Under certain model assumptions, the jet opening angle can be estimated from the afterglow \citep{1999ApJ...519L..17S,2001ApJ...562L..55F} or the prompt emission of LGRBs \citep{2016ApJ...818...18G}, with  mean reported values being roughly in the range of approximately 3 to 20 degrees (corresponding to $f_\mathrm{B}$ of 0.001-0.06). For our fiducial model we chose $f_\mathrm{B}=0.05$ and $\eta=0.01$. Different choices of $f_\mathrm{B}$, given the calibration, result in different LGRB rate densities as shown in Figure~\ref{fig:rate_uncertainties}. Lower $f_\mathrm{B}$ values lead to a suppression of the rates as the chance of seeing these systems are directly proportional to $f_\mathrm{B}$, while the contrary is true for larger $f_\mathrm{B}$ values.


\section{BBH and LGRB rate densities and detection rate}\label{sec:rates}

The BBH merger rate density $\mathcal{R}_\mathrm{BBHs}(z)$ is the number of BBHs mergers per comoving volume per year as a function of redshift. This quantity can be calculated \citep{2021A&A...647A.153B} by convolving the redshift- and metallicity-dependent star-formation rate (SFR) density with the synthetic BBH population obtained sampling initial binary distributions. To conduct this calculation, we assume a flat $\Lambda$CDM cosmology with $H_0 = 67.7~\mathrm{km\,s^{-1}\,Mpc^{-1}}$ and $\Omega_m=0.307$ \citep{2016A&A...594A..13P}.

We assume a modeled redshift- and metallicity-dependent star formation rate, $\mathrm{SFR}(z,\log_{10}(Z))$, from the TNG100 Illustris simulation \citep{2019ComAC...6....2N}. Illustris is a state-of-the-art large-scale cosmological simulation of the Universe. This model tracks the expansion of the Universe assuming a flat $\Lambda$CDM cosmology, the gravitational pull of baryonic and dark matter onto itself, the hydrodynamics of cosmic gas, as well as the formation of stars. The simulated comoving volume of $(100 \mathrm{Mpc})^3$ contains tens of thousands of galaxies captured in high detail. Illustris is calibrated to match the present-day ratio of the number of stars to dark matter for galaxies of all masses and the total amount of star formation in the universe as a function of time. Furthermore, the simulation also matches the galaxy stellar mass and luminosity functions.


The population synthesis predictions are performed in finite time bins of $\Delta t_i = 100~\mathrm{Myr}$ and log-metallicity bins $\Delta Z_j$. Each binary k with BH masses $m_{1,k}$ and $m_{2,k}$ is placed at redshift of formation $z_{\mathrm{f},i}$ corresponding to the center of $\Delta t_i$ and merging at redshift $z_{\mathrm{m},i,k}$ for its corresponding metallicity bin $\Delta Z_j$. The BBH rate density is given by the Monte Carlo sum \citep{2021A&A...647A.153B}
\begin{equation}
    \mathcal{R}_\mathrm{BBHs} (z_i) = \sum_{\Delta Z_j} \sum_{k} f_\mathrm{corr} \frac{\mathrm{fSFR}(z_{\mathrm{f},i}|\Delta Z_j)}{M_{\mathrm{sim,} \, \Delta Z_j}}
    \frac{4 \pi c \, D^2_\mathrm{c}(z_{\mathrm{m},i,k})}{\Delta V_\mathrm{c}(z_i)} \, \Delta t_i \, \, \, \mathrm{Gpc^{-3} yr^{-1}},
    \label{eq:RBBHs}
\end{equation}
where $M_{\mathrm{sim},\Delta Z_j}$ is the simulated mass per log-metallicity bin $\Delta Z_j$ and $f_\mathrm{corr}$ the normalization constant which converts the simulated mass to the total stellar population \citep{2020A&A...635A..97B}.
Here, 
$\mathrm{fSFR}(z|\Delta Z_j)=\int_{\Delta Z_j}\mathrm{SFR}(z,\log_{10}(Z)) \log_{10} Z$ is the fractional SFR density corresponding to the log-metallicity bin $\Delta Z_j$
and $\Delta V_\mathrm{c}(z_i)$ is the comoving volume shell corresponding to $\Delta t_i$, 
\begin{equation}
    \Delta V_\mathrm{c}(z_i) \equiv \int_{\Delta z_i} \frac{1}{1+z} \frac{\mathrm{d} V_\mathrm{c}}{\mathrm{d}z} \mathrm{d}z = \frac{4\pi c}{H_0} \int_{\Delta z_i} \frac{D_\mathrm{c}^2(z)}{E(z)(1+z)} \mathrm{d}z \, ,
\end{equation}
where, $\Delta z_i$ is the redshift interval corresponding to the formation time bin $\Delta t_i$, $D_c(z)=c/H_0 \int_0^z E(z')^{-1} dz$ is the comoving distance, $E(z)=\sqrt{\Omega_m(1+z)^3+\Omega_\Lambda}$ and $\Omega_\Lambda = 1 -\Omega_m$. 

A fraction of merging BBHs emit LGRBs at the compact object's formation, i.e., $z^l_{\mathrm{LGRB},i,k}$ where the dummy index $l=1,2$ indicates the first- or second-formed BH. In the case of CE and SMT channels, only the second-born tidally spun up BH can lead to a LGRB event, while for the CHE channel, we assume both stars can emit the LGRB at the same time $z^1_{\mathrm{LGRB},i,k}=z^2_{\mathrm{LGRB},i,k}$. We can therefore compute the LGRB rate density $R_\mathrm{LGRB}(z)$ by substituting $z_{\mathrm{LGRB},i,k}$ to $z_{m,i,k}$ in Eq.~(\ref{eq:RBBHs}). Accounting for beaming, we obtain the LGRB rate density visible to an observer as
\begin{equation}
    \mathcal{R}_\mathrm{LGRB}(z) = \sum_{\Delta Z_j} \sum_{k} f_\mathrm{B} f_\mathrm{corr} \frac{\mathrm{fSFR}(z_{\mathrm{f},i}|\Delta Z_j)}{M_{\mathrm{sim,} \, \Delta Z_j}}
    \frac{4 \pi c \, D^2_\mathrm{c}(z_{\mathrm{LGRB},i,k})}{\Delta V_\mathrm{c}(z_i)} \, \Delta t_i \, \, \, \mathrm{Gpc^{-3} yr^{-1}} \, .
    \label{eq:LGRBs}
\end{equation}

\begin{figure}[ht]
\centering
\subfloat{\includegraphics[width=0.7\linewidth]{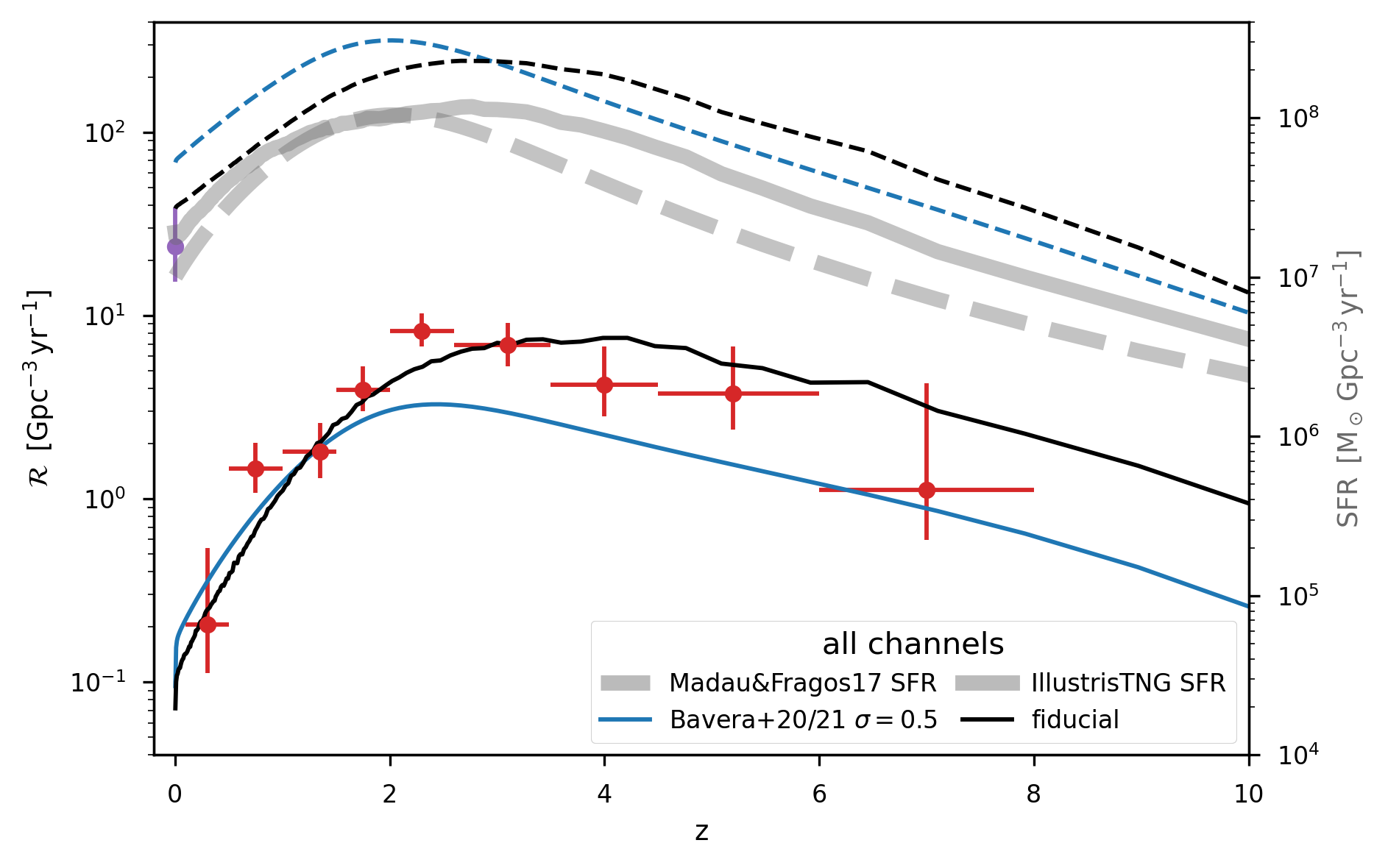}}
\caption{Modeled merging BBH (dashed lines) and luminous LGRB (solid lines) rate densities as a function of redshift for all channels combined. The figure illustrates model uncertainties given an alternative choice of SFR density \citep{2017ApJ...840...39M} (dashed gray line) and assuming metallicities follow a truncated log-normal metallicity with $\sigma = 0.5$ dex as in \citep{2020A&A...635A..97B,2021A&A...647A.153B}, in blue, versus the fiducial assumption of IllustrisTNG SFR density \citep{2019ComAC...6....2N} (solid gray line), in black. The fiducial luminous LGRB rate estimate assumes the beaming fraction $f_\mathrm{B} = 0.05$ and LGRB energy efficiency $\eta=0.0.01$, while the alternative model was calibrated against the empirical isotropic-equivalent energy to $f_\mathrm{B}=0.02$ and $\eta=0.002$.
}
\label{fig:SFR_uncertatinties}
\end{figure}

To highlight the uncertainties in the SFR density and metallicity distribution which might bias our rate estimate, we compare our results given the fiducial SFR density choice \citep[IllustrisTNG,][]{2019ComAC...6....2N} to an alternative SFR density \citep{2017ApJ...840...39M} assumed in previous works \citep{2020A&A...635A..97B,2021A&A...647A.153B} where it was assumed that metallicity follow a truncated log-normal distribution around the empirical mean of \citep{2017ApJ...840...39M} with $\sigma=0.5$ dex. In Figure~\ref{fig:SFR_uncertatinties}, we see that IllustrisTNG SFR density peaks at slightly higher redshift $z\in[2,3]$ compared to \citet{2017ApJ...840...39M} SFR density which peaks at $z=2$ while the latter shows a larger suppression at higher redshifts. Moreover, the alternative model predicts twice the fiducial BBH rate densities for $z<2$. The difference lies in the metallicity distribution which in the alternative model predicts more low metallicity systems compared to the IllustrisTNG metallicity distribution. This difference is due to the truncation of the log-normal distribution centered around the empirical mean which shifts the distribution towards lower metallicity systems and, hence, leads to an overproduction of merging BBH systems compared to IllustrisTNG.


The BBH detection rate $R_\mathrm{BBHs}$ is the number of BBH mergers observed per year by a gravitational-waves detector network. Similarly to the rate density calculation, we can calculate the BBH detection rate with the Monte Carlo sum \citep{2021A&A...647A.153B}
\begin{equation}
    R_\mathrm{BBHs} = \sum_{\Delta t_i, \Delta Z_j,k} w_{i,j,k} = \sum_{\Delta t_i} \sum_{\Delta Z_j} \sum_{k} p_{\mathrm{det},i,k} \, f_\mathrm{corr} \frac{\mathrm{fSFR}(z_{\mathrm{f},i}|\Delta Z_j)}{M_{\mathrm{sim,} \, \Delta Z_j}}
    4 \pi c \, D^2_\mathrm{c}(z_{\mathrm{m},i,k}) \, \Delta t_i \, \, \, \mathrm{yr}^{-1},
    \label{eq:w_BBH}
\end{equation}
where $w_{i,j,k}$ is the contribution of the BBH $k$ to the detection rate. Similarly to the rate density calculation, the binary k is placed at the time bin $\Delta t_i$ with center the redshift of formation $z_{\mathrm{f},i}$ and merging at $z_{\mathrm{m},i,k}$ for its corresponding metallicity bin $\Delta Z_j$. Here, $p_{\mathrm{det},i,k}\equiv p_\mathrm{det}(z_{\mathrm{m},i,k}, m_{1,k}, m_{2,k},\textbf{a}_{1,k},\textbf{a}_{1,k})$ is the detection probability which account for selection effects of the detector. Each BBH $k$ is characterised by the masses $m_{1,k}$ and $m_{2,k}$, and by the dimensionless spin vectors $\textbf{a}_{1,k}$ and $\textbf{a}_{2,k}$.
To compute $p_{\mathrm{det},i,k}$ \citep{2021A&A...647A.153B} we assume a three detector configuration with a network signal-to-noise ratio threshold of $12$ and ``mid-high/late-low'' sensitivity \citep{2018LRR....21....3A}, consistent with the third observing run of LIGO and Virgo detectors \citep{2021A&A...647A.153B,2021ApJ...910..152Z}.

The normalised weight $\tilde{w}_{i,j,k}=w_{i,j,k}/\sum_{\Delta t_{i'}, \Delta Z_{j'}, {k'}} w_{i',j',k'}$ is used to generate the gravitational-waves observable distributions of the detected BBH modeled population in the left panel of Figure~\ref{fig:model_vs_GWTC2}. To generate the underlying (intrinsic) BBH merging distribution in Figure~\ref{fig:model_vs_GWTC2}, i.e. what a detector on Earth with infinite sensitivity would observe, we weight the modeled population with $\tilde{w}_{i,j,k}^\mathrm{intrinsic} = \tilde{w}_{i,j,k}(p_{\mathrm{\det},i,k}=1)$. Finally the intrinsic distribution of BBH mergers associated with luminous LGRBs shown in the right panel of Figure~\ref{fig:model_vs_GWTC2} is obtain by weighting the modeled population as
\begin{equation}
    \tilde{w}_{i,j,k}^\mathrm{intrinsic,LGRB} = \begin{cases} 
    \tilde{w}_{i,j,k}^\mathrm{intrinsic} , \, E^\mathrm{iso}_\mathrm{LGRB} > 10^{51} \mathrm{erg} \\ 
    0 , \,\,\, \mathrm{else}
    \end{cases} .
    \label{eq:w_LGRB}
\end{equation}

\section{Luminous LGRB evidence in GWTC-2}\label{sec:LGRB_prob}

The probability of a gravitational-wave event $\textbf{x}$ to have emitted a luminous LGRB, given our model, is calculated as
\begin{equation}
    \begin{split}
    p_\mathrm{LGRB}(\textbf{x}) &= \int_{-1}^{1} \int_{0\,\Msun}^{100\,\Msun} f_\mathrm{GRB}(\chieff,\mchirp) \times p(\chieff,\mchirp|\textbf{x}) \, \mathrm{d}\chieff \, \mathrm{d}\mchirp = \\
    & \approx \sum_l \sum_m f_\mathrm{LGRB}^{l,m} p(\Delta \chieff^l, \Delta \mchirp^m) \Delta \chieff \Delta\mchirp \, ,
    \end{split}
\end{equation}
where we approximated the integrals with a Riemann sum over the finite $l$- and $m$-bins of size $\Delta \chieff=0.05$ and $\Delta \mchirp=2\,\Msun$, respectively. The gravitational-waves events' posterior probability density $p(\chieff,\mchirp|\textbf{x})$ is discretised and calculated at the center of each 2D bin ($\Delta \chieff^l$,$\Delta \mchirp^m$). Here, $f_\mathrm{LGRB}$, is the probability density of an event with ($\chieff$,$\mchirp$) to have emitted a luminous LGRB at BBH formation. We approximate this probability, given our model, over the finite bins $\Delta \chieff^l$ and $\Delta \mchirp^m$ as
\begin{equation}
    f_\mathrm{LGRB}^{l,m} \equiv  f_\mathrm{LGRB}(\Delta \chieff^l, \Delta \mchirp^m) = \frac{\sum_{\Delta t_i, \Delta Z_j, k} w^\mathrm{intrinsic,LGRB}_{i,j,k} (\Delta \chieff^l, \Delta \mchirp^m)}{\sum_{\Delta t_{i'}, \Delta Z_{j'}, {k'}} w^\mathrm{intrinsic}_{i',j',k'}(\Delta \chieff^l, \Delta \mchirp^m)} \, ,
\end{equation}
where $w^\mathrm{intrinsic}_{i,j,k}$ is the weight contribution of each binary to the intrinsic detection rate and $w^\mathrm{intrinsic,LGRB}_{i,j,k}$ is conditioned against the luminous LGRB criteria similar to Eq.~(\ref{eq:w_LGRB}).

The probability $p_\mathrm{LGRB}$ of each event in GWTC-2 is summarised in Table~\ref{tab:LGRB_prob}, where we also report as a reference the median $\chieff$ and $\mchirp$ of each event. 

\begin{table}
\centering
\resizebox{0.52\columnwidth}{!}{%
\begin{tabular}{ |c|c|c|c| } 
\hline
& emitted & &  \\
EVENT & LGRB  & ${<}\chieff{>}$ & ${<}\mchirp{>}$ \\
& chance in \% & & $[\mathrm{M}_\odot]$ \\
\hline
GW190517\_055101 & 86.85 & 0.52 & 26.6 \\
GW190719\_215514 & 59.82 & 0.31 & 23.4 \\
GW190412 & 37.88 & 0.25 & 13.3 \\
GW170729 & 28.37 & 0.37 & 35.4 \\
GW190828\_063405 & 26.93 & 0.19 & 25.0 \\
GW190527\_092055 & 19.00 & 0.11 & 24.3 \\
GW190513\_205428 & 18.89 & 0.11 & 21.6 \\
GW190727\_060333 & 15.36 & 0.11 & 28.7 \\
GW151012 & 13.26 & 0.05 & 15.2 \\
GW190424\_180648 & 10.29 & 0.13 & 31.1 \\
GW190620\_030421 & 9.27 & 0.33 & 38.2 \\
GW170823 & 7.68 & 0.09 & 29.2 \\
GW190731\_140936 & 6.39 & 0.06 & 29.6 \\
GW190413\_052954 & 5.94 & -0.01 & 24.6 \\
GW170809 & 5.57 & 0.08 & 24.9 \\
GW190828\_065509 & 4.20 & 0.08 & 13.3 \\
GW190930\_133541 & 4.15 & 0.14 & 8.5 \\
GW190630\_185205 & 3.44 & 0.09 & 24.9 \\
GW190915\_235702 & 2.96 & 0.02 & 25.3 \\
GW190803\_022701 & 2.54 & -0.03 & 27.3 \\
GW190909\_114149 & 2.04 & -0.06 & 30.6 \\
GW151226 & 2.01 & 0.18 & 8.9 \\
GW190706\_222641 & 1.82 & 0.28 & 42.8 \\
GW190413\_134308 & 1.62 & -0.04 & 32.9 \\
GW170814 & 1.38 & 0.07 & 24.1 \\
GW190929\_012149 & 1.00 & 0.01 & 35.8 \\
GW190519\_153544 & 0.79 & 0.31 & 44.6 \\
GW190512\_180714 & 0.62 & 0.03 & 14.6 \\
GW190421\_213856 & 0.55 & -0.06 & 31.2 \\
GW190728\_064510 & 0.49 & 0.12 & 8.6 \\
GW170104 & 0.47 & -0.04 & 21.4 \\
GW190503\_185404 & 0.44 & -0.03 & 30.2 \\
GW190521\_074359 & 0.41 & 0.09 & 32.1 \\
GW190720\_000836 & 0.34 & 0.18 & 8.9 \\
GW190514\_065416 & 0.25 & -0.19 & 28.7 \\
GW170818 & 0.18 & -0.09 & 26.6 \\
GW190910\_112807 & 0.15 & 0.02 & 34.3 \\
GW190924\_021846 & 0.09 & 0.03 & 5.8 \\
GW170608 & 0.07 & 0.03 & 7.9 \\
GW190408\_181802 & 0.07 & -0.03 & 18.3 \\
GW190708\_232457 & 0.07 & 0.02 & 13.2 \\
GW190707\_093326 & 0.00 & -0.05 & 8.5 \\
GW150914 & 0.00 & -0.01 & 28.6 \\
GW190602\_175927 & 0.00 & 0.07 & 49.2 \\
GW190521 & 0.00 & 0.03 & 69.2 \\
GW190701\_203306 & 0.00 & -0.07 & 40.3 \\
\hline
\hline
CUMULATIVE & 383.66 & & \\
\hline

\end{tabular}
}
\caption{Probabilities of each BBH event in GWTC-2 to have  emitted a luminous LGRB, $E_\mathrm{LGRB}^\mathrm{iso} > 10^{51} \, \mathrm{erg}$, at the formation of the BBH system. For comparison, we report the median $\chieff$ and $\mchirp$ for each event. The expected number of GWTC-2 events that had emitted a luminous LGRB is $\approx$4 out of 46.
}
\label{tab:LGRB_prob}
\end{table}

\section{Metallicity of LGRB progenitors}\label{sec:metallicity}

The maximal ZAMS metallicity of LGRB progenitors in our models is primarily  dictated by the interplay of tides and Wolf-Rayet stellar winds \citep{2000A&A...360..227N}, which is the dominant phase of stellar wind mass loss and is taken to scale with metallicity as $\propto (Z/Z_\odot)^{0.85}$ \citep{2001A&A...369..574V}. In our model, this threshold is at $Z_\mathrm{max}\approx0.2\,Z_\odot$, where we adopt $Z_\odot=0.017$ \citep{1996ASPC...99..117G}. As shown in Figure~\ref{fig:metallicity}, this corresponds to the lower 16\% bound of the metallicity distribution of newly formed stars at $z=2$ in the IllustrisTNG simulation, which we use as input in our models. In the same figure we compare the progenitors' metallicities of modeled LGRBs to the sub-sample of the SHOALS LGRBs with 45 identified host galaxies which have measured metallicities for $z<2.5$ \citep{2019arXiv190402673G}. We have translated the reported $12+\log_{10}(\mathrm{O/H})$ to $\mathrm{[Fe/H]}$ using an empirical relation between $\mathrm{[O/Fe]}$ and $\mathrm{[Fe/H]}$ \citep{2017MNRAS.466.4403N} and took the solar reference as $\mathrm{[O/H]}_\mathrm{ref}=8.83$ \citep{1998SSRv...85..161G}. Explicitly, we numerically solve the equation $ [\mathrm{Fe/H}] = [\mathrm{O/H}] - [\mathrm{O/Fe}]([\mathrm{Fe/H}])$
with respect to $[\mathrm{Fe/H}]$ where $[\mathrm{O/H}] = 12+\log_{10}\left(\mathrm{O/H}\right) - \mathrm{[O/H]}_\mathrm{ref}$ and \citep[see Eq.~(5) in][]{2017MNRAS.466.4403N}
\begin{equation}
    [\mathrm{O/Fe}]\left([\mathrm{Fe/H}]\right) = \begin{cases}
    +0.5, \, & -2.5 < [\mathrm{Fe/H}] \leq -1 \\
    -0.5 \times [\mathrm{Fe/H}], \, & -1 < [\mathrm{Fe/H}] \leq 0.5 \\
    -0.25, \, & [\mathrm{Fe/H}] > 0.5 \, .
    \end{cases}
\end{equation}
Typical values of $\mathrm{[O/Fe]}$ increase as $\mathrm{[Fe/H]}$ decreases due to the increased influence of Type II supernovae over Type Ia at lower metallicities. At face value, we find that 40\% of the observed LGRB host galaxies have metallicities lower than $Z_\mathrm{max}$. However, when taking into account possible systematic uncertainties in the calibration of different metallicity measurement methods, we find that our model can be consistent between 18 and 86\% of all observed luminous LGRBs, cf. Figure~\ref{fig:metallicity}. These uncertainties can be as high as $\pm 0.35\,\mathrm{dex}$ on the measured abundance $\log_{10}(\mathrm{O/H})$ \citep{2008ApJ...681.1183K}, where \citep{2019arXiv190402673G} determined metallicities using the $R_{23}$ diagnostic scale of \citet{2004ApJ...617..240K} which are skewed towards larger values with respect to other calibration methods \citep[cf. Figure~2 of][]{2008ApJ...681.1183K}.

\end{document}